\documentclass[journal=jpclcd,manuscript=letter]{achemso}
\setkeys{acs}{maxauthors=0,articletitle=true}

\usepackage{achemso}
\usepackage{graphics}
\usepackage{amssymb,amsfonts}
\usepackage{graphicx}
\usepackage[table]{xcolor}
\usepackage{multirow}
\usepackage{caption}
\usepackage{subcaption}
\usepackage{booktabs}
\usepackage{colortbl}
\usepackage{amsmath}
\usepackage{amsopn}
\usepackage{siunitx}
\usepackage{bm}
\usepackage{color}
\usepackage{array}
\usepackage{lscape}
\usepackage{mciteplus}
\usepackage[version=3]{mhchem}
\usepackage{ulem}
\usepackage{listings}
\usepackage{enumerate}
\usepackage{lmodern}

\SectionNumbersOn

\author{Janus J. Eriksen}
\email{janus.eriksen@bristol.ac.uk}
\affiliation[University of Bristol]
{School of Chemistry, University of Bristol, Cantock's Close, Bristol BS8 1TS, United Kingdom}
\author{J{\"u}rgen Gauss}
\email{gauss@uni-mainz.de}
\affiliation[Johannes Gutenberg-Universit\"at Mainz]
{Institut f\"ur Physikalische Chemie, Johannes Gutenberg-Universit\"at Mainz, Duesbergweg 10-14, 55128 Mainz, Germany}

\title[TITLE]{Generalized Many-Body Expanded Full Configuration Interaction Theory}

\begin{tocentry}
\hspace{-2.8cm}
\includegraphics[scale=0.3]{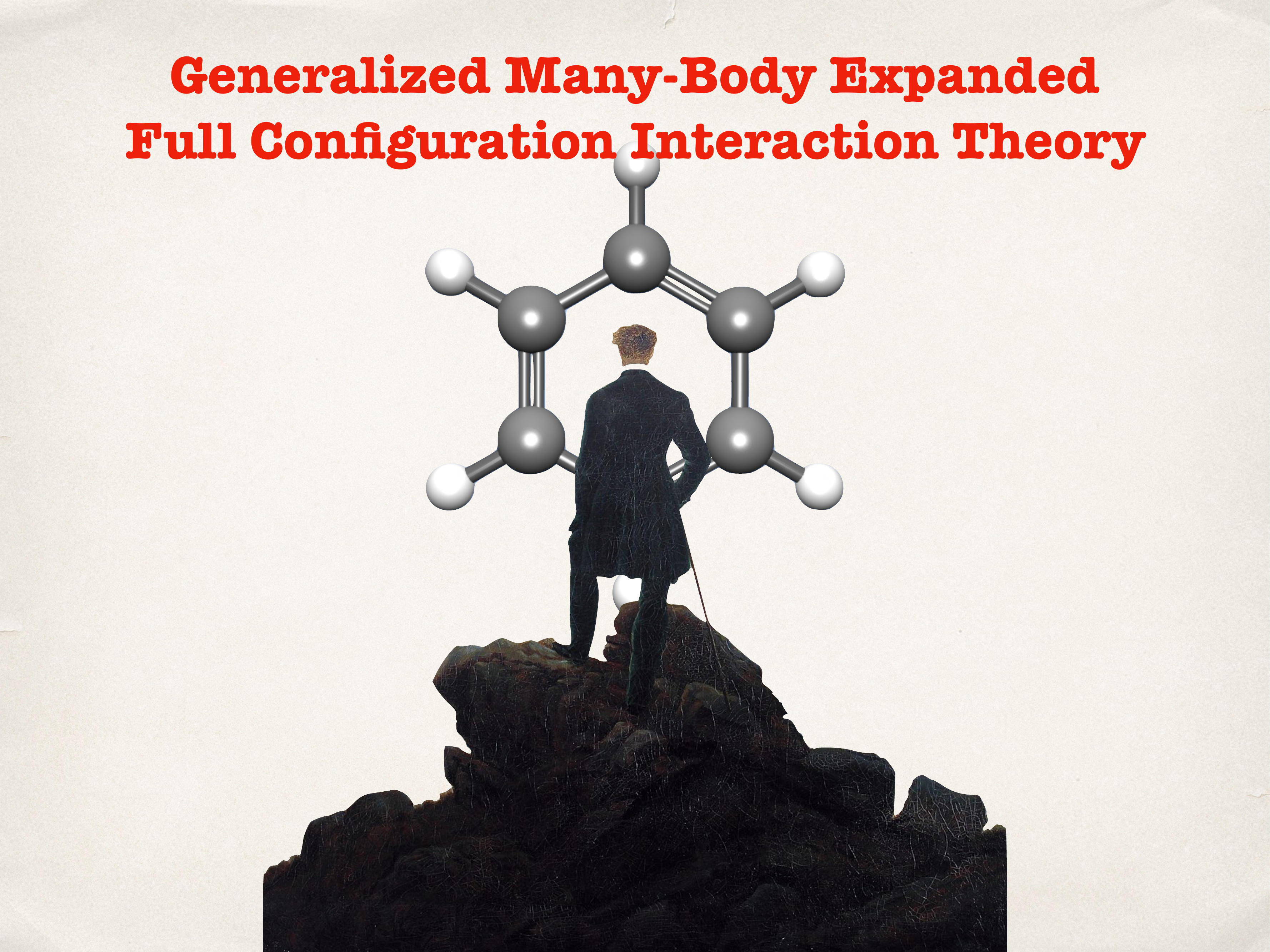}
\end{tocentry}

\begin{document}

%
%
\begin{abstract}
Facilitated by a rigorous partitioning of a molecular system's orbital basis into two fundamental subspaces---a reference and an expansion space, both with orbitals of unspecified occupancy---we generalize our recently introduced many-body expanded full configuration interaction (MBE-FCI) method to allow for electron-rich model and molecular systems dominated by both weak and strong correlation to be addressed. By employing minimal or even empty reference spaces, we show through calculations on the 1-dimensional Hubbard model with up to 46 lattice sites, the chromium dimer, and the benzene molecule how near-exact results may be obtained in a entirely unbiased manner for chemical and physical problems of not only academic, but also applied chemical interest. Given the massive parallelism and overall accuracy of the resulting method, we argue that generalized MBE-FCI theory possesses an immense potential to yield near-exact correlation energies for molecular systems of unprecedented size, composition, and complexity in the years to come.
\end{abstract}

\newpage

%
%

In electronic structure theory, the motion of the electrons of a system is governed by the time-independent electronic Schr{\"o}dinger equation. In practice, this fundamental equation is exceedingly difficult, if not impossible to solve as its wave function solutions are composite functions of the simultaneously correlated motion of all involved electrons. Analytical solutions for anything but simple one-electron systems are thus not possible, which is why approximations in finite one-electron basis sets to the exact full configuration interaction (FCI) wave function must be invoked. As the realization of FCI scales exponentially with respect to both the number of electrons, $N_{\text{elec}}$, and the number of basis functions, $M_{\text{orb}}$, the majority of pragmatic methods compromise accuracy against the FCI solution by explicitly truncating the wave function at a reduced computational scaling.~\cite{white_dmrg_prl_1992,white_dmrg_prb_1993,white_martin_dmrg_jcp_1999,chan_head_gordon_dmrg_jcp_2002,booth_alavi_fciqmc_jcp_2009,cleland_booth_alavi_jcp_2010,malrieu_selected_ci_jcp_1973,holmes_umrigar_heat_bath_ci_jctc_2016,sharma_umrigar_heat_bath_ci_jctc_2017,li_sharma_umrigar_heat_bath_ci_jcp_2018,fales_koch_martinez_rrfci_jctc_2018,liu_hoffmann_ici_jctc_2016,tubman_whaley_selected_ci_jcp_2016,loos_selected_ci_jcp_2018,schriber_evangelista_selected_ci_jcp_2016,schriber_evangelista_adaptive_ci_jctc_2017,lu_coord_descent_fci_jctc_2019,berkelbach_fci_fri_jctc_2019,blunt_sci_fciqmc_arxiv_2019}. However, progress has also been made in approximating the properties associated with the FCI solution directly. As a pertinent example, the present authors have recently proposed the many-body expanded full configuration interaction~\cite{eriksen_mbe_fci_jpcl_2017,eriksen_mbe_fci_weak_corr_jctc_2018,eriksen_mbe_fci_strong_corr_jctc_2019} (MBE-FCI) method in which FCI is decomposed by means of a many-body expansion (MBE) in a basis of molecular orbitals (MOs).

In the MBE-FCI method, one begins by enforcing a partitioning of the complete set of spatial MOs into a {\textit{reference}} and an {\textit{expansion}} space. Here, the reference space constitutes the subspace of MOs which are always included in all constituent CASCI calculations of the MBE-FCI method, which themselves in addition include an increasing number of MOs from the expansion space (augmentation of the reference space by single MOs at order $1$, all unique pairs at order $2$, etc.). An expansion in the latter space hence serves to recover the residual correlation missing from a CASCI calculation in the reference space, $E_{\text{ref}}$. Removing any restrictions on the composition of the reference space, that is, allowing for it to be filled by MOs from both the occupied and virtual subspaces (of sizes $M_{\text{occ}}$ and $M_{\text{virt}}$, respectively), the decomposition of the FCI correlation energy is formally written as
\begin{align}
E_{\text{FCI}} &= E_{\text{ref}} + \sum_{p}\epsilon_{p} + \sum_{p<q}\Delta\epsilon_{pq} + \sum_{p<q<r}\Delta\epsilon_{pqr} + \ldots \nonumber \\
&\equiv E_{\text{ref}} + E^{(1)} + E^{(2)} + E^{(3)} + \ldots + E^{(M_{\text{exp}})} \label{mbe_eq}
\end{align}
where the MOs of the expansion space (of size $M_{\text{exp}}$) of unspecified occupancy are labelled by indices $\{p,q,r,s,\ldots\}$, and $\epsilon_{p}$ designates the energy of a CASCI calculation in the composite space of orbital $p$ and all of the reference space MOs. The increments of order $n$, $\Delta\epsilon_{[\Omega]_{n}}$, are recursively defined through the following relation for a general tuple of $n$ MOs, $[\Omega]_{n}$, as
\begin{align}
\Delta\epsilon_{[\Omega]_{n}} = \epsilon_{[\Omega]_{n}} - \sum_{p \in S_1[\Omega]_{n}}\epsilon_{p} - \sum_{pq \in S_2[\Omega]_{n}}\Delta\epsilon_{pq} - \ldots - \sum_{pqrs\cdots \in S_{n-1}[\Omega]_{n}}\Delta\epsilon_{pqrs\cdots} \label{increment_eq}
\end{align}
where the action of $S_{m}$ onto $[\Omega]_{n}$ is to construct all possible unique subtuples of order $m$ ($1 \leq m < n$) and $\epsilon_{[\Omega]_{n}}$ is defined on par with $\epsilon_{p}$ in Eq. \ref{mbe_eq}.
\begin{figure}[ht]
\begin{center}
\includegraphics[scale=0.38]{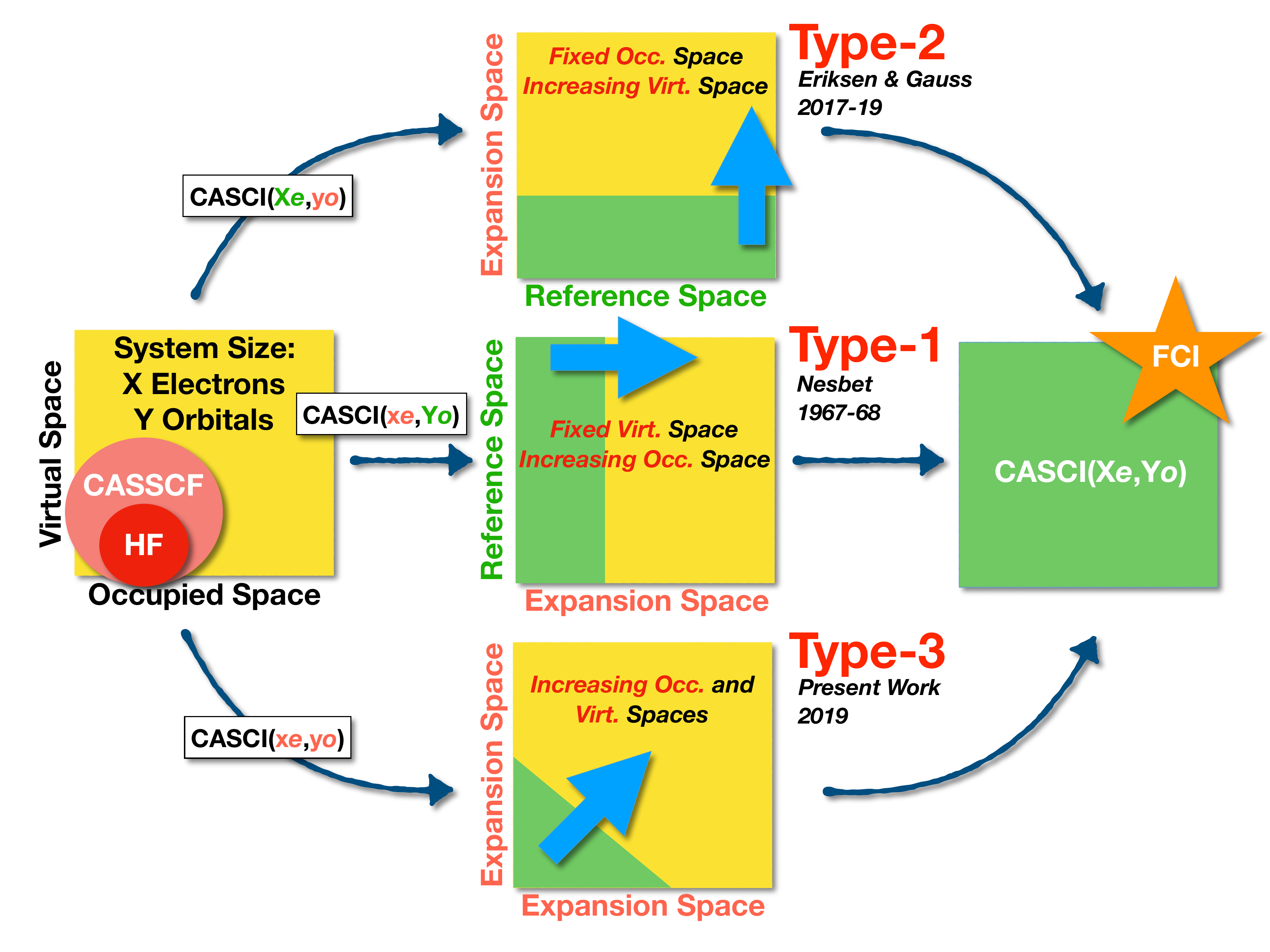}
\caption{Schematic representation of the various manners ({\textbf{Types 1}}, {\textbf{2}}, and {\textbf{3}}) in which to decompose the FCI solution by means of MBE-FCI.}
\label{mbe_fci_scheme_fig}
\end{center}
\vspace{-0.6cm}
\end{figure}

In the original Bethe-Goldstone method~\cite{nesbet_phys_rev_1967_1,nesbet_phys_rev_1967_2,nesbet_phys_rev_1968}, Nesbet proposed to perform an MBE in the basis of the occupied orbitals of a system, i.e., correlating an increasing amount of electron pairs among each other while keeping the available virtual correlation space fixed at all times. This is what is designated as {\textbf{Type--1}} MBE-FCI in Figure \ref{mbe_fci_scheme_fig}. Given that an MBE in a basis of occupied orbitals converges rapidly, say, at a maximum order $k$ (corresponding to correlation among $2k$ electrons) where $k \ll M_{\text{occ}}$, the MBE-FCI method in its {\textbf{Type--1}} form holds promise of being able to be applied to electron-rich systems. However, since the complete virtual space remains available to excitations out of the occupied MOs of the expansion space, an unfavorable factorial scaling of the individual CASCI calculations with $M_{\text{orb}}$ effectively restricts {\textbf{Type--1}} MBE-FCI to very small basis sets. In an attempt at remedying this bottleneck, Zimmerman~\cite{zimmerman_ifci_jcp_2017_1,zimmerman_ifci_jcp_2017_2,zimmerman_ifci_jpca_2017,zimmerman_ifci_jcp_2019} recently proposed to explicitly truncate the extent of the virtual space based on orbital occupancy measures. While this makes {\textbf{Type--1}} MBE-FCI calculations in medium-sized basis sets feasible, one remains forced to rely on low-order approximations to $E_{\text{FCI}}$, the accuracy of which is generally hard to assess {\textit{a priori}}.

In {\textbf{Type--2}} MBE-FCI~\cite{eriksen_mbe_fci_jpcl_2017,eriksen_mbe_fci_weak_corr_jctc_2018,eriksen_mbe_fci_strong_corr_jctc_2019}, the foundations behind the {\textbf{Type--1}} variant were inverted by the present authors through the use of particle-hole symmetry, cf. Figure \ref{mbe_fci_scheme_fig}. By defining the complete set of occupied orbitals (in addition to potentially the virtual orbitals belonging to an active space~\cite{eriksen_mbe_fci_strong_corr_jctc_2019}) as the reference space and all (inactive) virtual orbitals as constituting the expansion space, it has been demonstrated how both weakly and strongly correlated systems may be treated by MBE-FCI using extended basis sets. However, as all electrons are correlated in all of the involved CASCI calculations, which themselves involve an increasingly large number of orbitals throughout the MBE, the steep scaling with $N_{\text{elec}}$ prevents {\textbf{Type--2}} MBE-FCI from being applied to systems with more than a modest number of electrons.

In the present work, we introduce the concept of minimal or even empty {\textit{vacuum}} reference spaces in order to simultaneously remove the remaining scaling walls of {\textbf{Type--1}} and {\textbf{Type--2}} MBE-FCI, cf. Figure \ref{mbe_fci_scheme_fig}. In the new {\textbf{Type--3}} MBE-FCI method, assuming a vacuum reference space, all MOs are thus treated as members of a collective expansion space, which may be viewed as offering the most general take on MBE-FCI. In this limit, Eq. \ref{mbe_eq} becomes
\begin{align}
E_{\text{FCI}} &= \sum_{p<q}\epsilon_{pq} + \sum_{p<q<r}\Delta\epsilon_{pqr} + \sum_{p<q<r<s}\Delta\epsilon_{pqrs} + \ldots \label{mbe_eq_vacuum}
\end{align}
where only tuples that index both occupied $\{i,j,\ldots\}$ and virtual $\{a,b,\ldots\}$ MOs give rise to non-vanishing increments ($\epsilon_{ia}$, $\Delta\epsilon_{ija}$, $\Delta\epsilon_{iab}$, etc.) and the expansion hence starts at order $2$.

As in {\textbf{Type--2}} MBE-FCI, {\it{child tuples}} at order $k+1$ are spawned from the complete set of $k$-order {\it{parent tuples}} subject to a screening threshold. For a given parent tuple, all subtuples of order $k-1$ are generated and augmented by a candidate child orbital, and a given child tuple is then spawned only if the absolute value of any of the increments that correspond to this set of $k$th-order tuples is greater than a numerical energy threshold~\cite{eriksen_mbe_fci_weak_corr_jctc_2018}. However, because the MBE-FCI expansion space now consists of both occupied and virtual MOs, and the tuples at any given order in general involve a varying number of these, the manner in which thresholds for individual tuples (tup) are calculated have been modified according to the following expression
\begin{align}
T_{\text{tup}} \ (\text{in} \ E_{\text{H}}) \equiv 
\left\{\begin{array}{ll}
0.0 & \text{if} \ \ M_{\text{tup}} < M_{\text{init}} \ \phantom{.} \\
T_{\text{init}} \cdot a^{M_{\text{tup}}-M_{\text{init}}} & \text{if} \ \ M_{\text{init}} \leq M_{\text{tup}}
\end{array}\right. \label{thres_eq}
\end{align}
where $M_{\text{tup}} =  \max (M_{\text{occ,tup}}, M_{\text{virt,tup}})$ and $M_{\text{occ,tup}}, M_{\text{virt,tup}}$ denote the numbers of occupied and virtual MOs of a given tuple, respectively. The screening thresholds are hence defined in a manner similar to the original definition in Ref. \citenum{eriksen_mbe_fci_jpcl_2017}, and all other parameters remain the same as in {\textbf{Type--2}} MBE-FCI, i.e., $T_{\text{init}} = \num{1.0e-10}$ $E_{\text{H}}$, $M_{\text{init}} = 3$, and $a \geq 1.0$, in practice leaving the latter relaxation factor ($a$) as the only adjustable MBE-FCI parameter.

Common to all of the results to follow, these were obtained using vacuum expansion references in order to emphasize unbiased versatility across different types of electron correlation and molecular size. Furthermore, a relatively aggressive screening threshold of $a=5.0$ was used throughout (unless otherwise noted), as the focus in the present work will not be on producing rigorous benchmark results but rather to illustrate the applicability enhancement of generalized MBE-FCI theory. The code used to perform the MBE-FCI calculations is our Python-based, open-source {\texttt{PyMBE}} code~\cite{pymbe}, which utilizes the {\texttt{PySCF}} program~\cite{pyscf_paper,pyscf_prog} for all electronic structure kernels and the {\texttt{MPI4Python}} module~\cite{mpi4py_1,mpi4py_2,mpi4py_3} for parallel MPI communication.

In extending the {\texttt{PyMBE}} code to be able to perform {\textbf{Type--3}} MBE-FCI calculations, we have further optimized the code, in particular the general memory handling and footprint of the 1- and 2-electron repulsion integrals as well as all involved intermediates and results. This has been achieved by pursuing an MPI+MPI approach to hybrid programming~\cite{hoefler_mpi_mpi_computing_2013} where the {\texttt{MPI$\_$Win$\_$allocate$\_$shared}} function is used as a departure from the standard abstract and distributed memory model of MPI. Essentially, the underlying memory organization on a given computer node is exposed to MPI, which allows {\texttt{PyMBE}} to bypass the expensive and somewhat convoluted MPI-3 one-sided operations and instead use shared memory directly between the processes on said node. Thus, MPI is employed in {\texttt{PyMBE}} in entirely different manners across and within individual nodes. For an MBE-FCI calculation on $n$ nodes each with $N$ cores, the parallel model consists of a single global master and $nN-1$ global slaves. Among the slaves, $n-1$ of these additionally serve as local masters that pass messages to the global master via distributed MPI, with each sharing a window to the physical memory address space on their respective node with $N-1$ local slaves over dedicated communicators.

In terms of computational resources, all results have been obtained in parallel on either of two systems: (i) a single Intel Xeon Broadwell E$5$--$2699$ $\text{v}4$ node with a total of $44$ cores {@} $2.20$ GHz and $768$ GB of global memory, that is, on readily available commodity hardware, or (ii) the Galileo system at CINECA in Bologna, Italy, which is equipped with 360 Intel Xeon E5-2697 v4 nodes, each comprising 36 cores @ 2.3 GHz and 128 GB of global memory. Tabulated data are collected in the Supporting Information (SI).

\begin{figure}[ht]
\begin{center}
\includegraphics[scale=0.9]{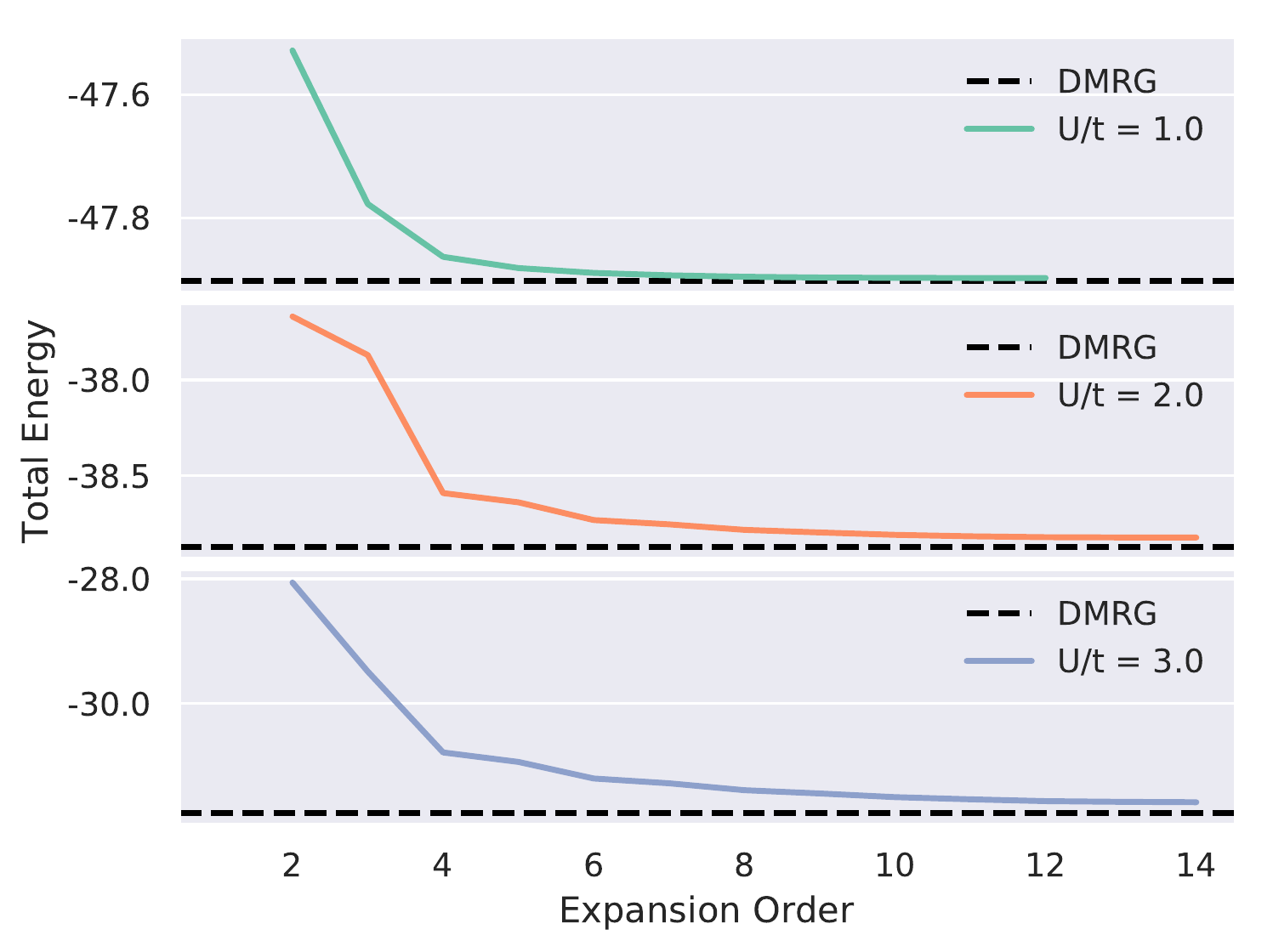}
\caption{{\textbf{Type--3}} MBE-FCI for the 46-site 1-dimensional Hubbard model. Results in $E_{\text{H}}$.}
\label{hubbard_1d_fig}
\end{center}
\vspace{-0.6cm}
\end{figure}
As an initial test of {\textbf{Type--3}} MBE-FCI, Figure \ref{hubbard_1d_fig} presents results for the half-filled 1-dimensional Hubbard model~\cite{hubbard_hubbard_model_prsa_1963,gutzwiller_hubbard_model_prl_1963,kanamori_hubbard_model_ptp_1963,simons_collab_hubbard_prx_2015}, which allows for a smooth model transition from weak to strong correlation and validation against analytical results for the thermodynamic limit (TDL) at arbitrary $U/t$ value~\cite{lieb_wu_hubbard_model_prl_1968} ($t$ and $U$ are the hopping and interaction terms, respectively, of the Hubbard Hamiltonian). In its current form, MBE-FCI remains inapplicable in the treatment of genuine $N$-body electron correlation, and for this reason we will limit ourselves to model Hamiltonians defined by $U/t\leq3$, which still allows for testing the method in correlation regimes of chemical interest. The results of Figure \ref{hubbard_1d_fig} for 46 lattice sites have all been obtained using a basis of modified Pipek-Mezey~\cite{pipek_mezey_jcp_1989} (PM) localized orbitals~\cite{Note-1} (results for the 22- and 34-site models are presented as Figures S1 and S2 in the SI). As is clear from the results in Figure \ref{hubbard_1d_fig}, our {\textbf{Type--3}} MBE-FCI results generally converge toward the exact solutions, here represented by corresponding density matrix renormalization group (DMRG) results~\cite{Note-2}. For the larger and more strongly correlated lattice problems, i.e., the 34- and 46-site models with $U/t = 3.0$ for which the total correlation energies amount to a full $-5.6$ and $-7.5$ $E_{\text{H}}$, respectively, slightly tighter and less aggressive screening thresholds are needed in order to achieve the same accuracy as is met for the corresponding weakly correlated lattice problems of similar size, i.e., total errors of $\mathcal{O}(10^{-4}$ $E_{\text{H}}$/site), cf. Table S2 in the SI. However, we stress once again that the production of rigorous benchmark results is not our objective in the present Letter. Across all of the calculations in Figure \ref{hubbard_1d_fig}, the largest CASCI calculation involved a total of $1.18\cdot10^{7}$ determinants, which is in stark contrast to the more than $10^{25}$ determinants a hypothetical FCI calculation for the 46-site Hubbard model would comprise.

\begin{figure}[ht]
\begin{center}
\includegraphics[scale=0.8]{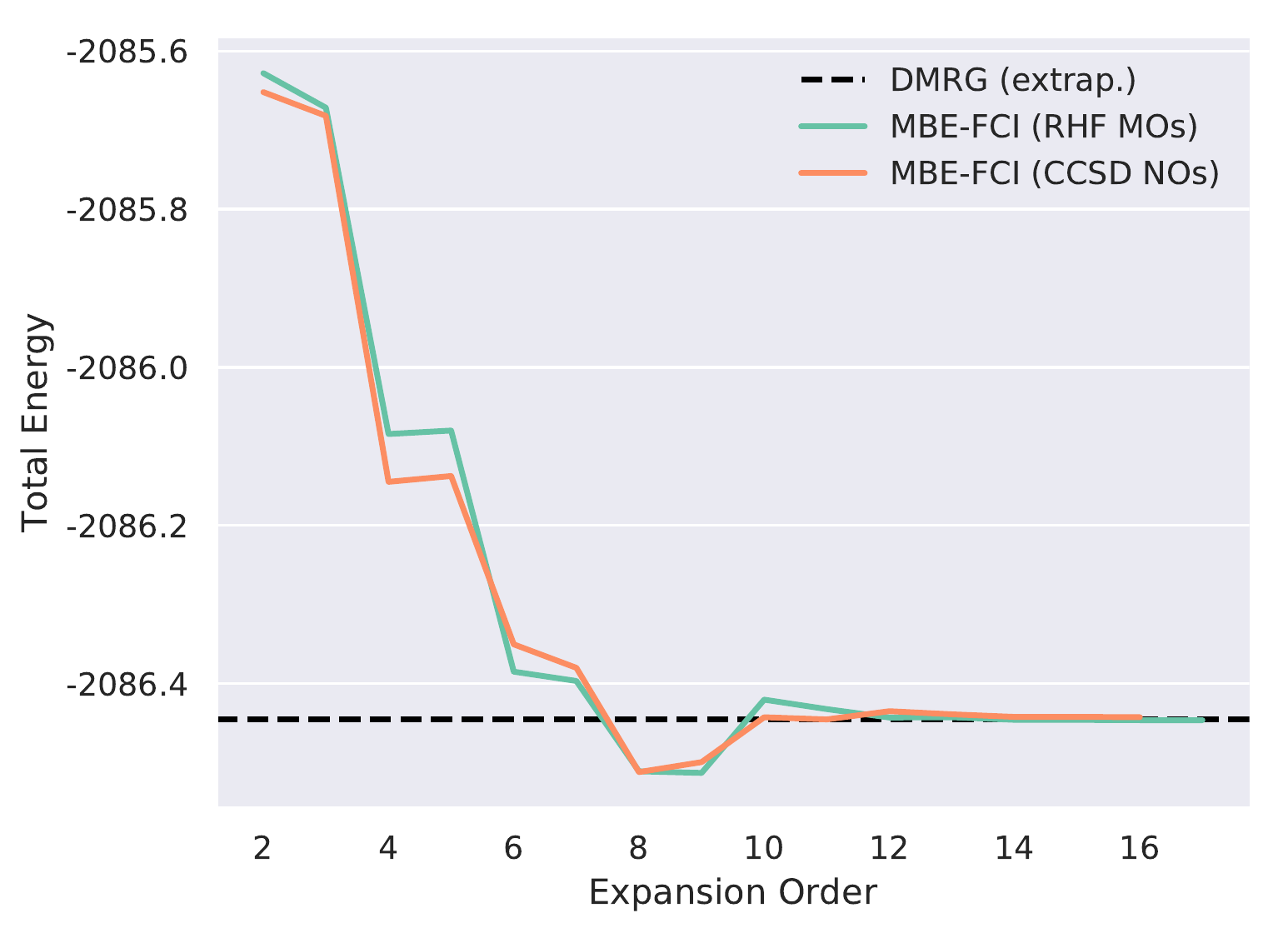}
\caption{{\textbf{Type--3}} MBE-FCI for Cr$_2$/Ahlrichs-SV ($r_e = 1.5$ {\AA}). Results in $E_{\text{H}}$.}
\label{cr2_dz_fig}
\end{center}
\vspace{-0.6cm}
\end{figure}
In Figure \ref{cr2_dz_fig}, we next present {\textbf{Type--3}} MBE-FCI results for the chromium dimer (Cr$_2$) in an Ahlrichs-SV basis set~\cite{ahrichs_sv}, which constitutes a favoured calibration example of a strongly correlated molecular system~\cite{roos_cr2_cccc_2003,vancoillie_malmqvist_veryazov_cr2_jctc_2016}. In the absence of spatial locality, we have here used two alternative MO bases, namely canonical restricted HF (RHF) MOs and coupled cluster singles and doubles~\cite{ccsd_paper_1_jcp_1982} (CCSD) natural orbitals (NOs), and we have correlated all 48 electrons in the full space of 42 MOs. Furthermore, our $\pi$-pruning prescreening filter of Ref. \citenum{eriksen_mbe_fci_strong_corr_jctc_2019} has been used in order to guarantee convergence onto the ground state of ${^{1}}\Sigma^{+}_g$ symmetry. In contrast to the Hubbard example in Figure \ref{hubbard_1d_fig}, no exact reference energy is available for Cr$_2$, even in the modest basis set used here, and we hence compare our results to a {\textit{de facto}} standard from the literature, which is an elaborately extrapolated DMRG result obtained by Chan and co-workers with a conservative error bar of $\pm0.32$ m$E_{\text{H}}$~\cite{chan_dmrg_2015}. As is visible from the expansion profiles in Figure \ref{cr2_dz_fig}, (i) both expansions converge onto the correct result and (ii) the advantages of optimized orbitals (NOs) are obvious as a more rapid convergence is achieved. However, the largest CASCI calculations at the final order of the expansion using NOs still comprised $2.07\cdot10^{7}$ determinants (with $D_{2\text{h}}$ symmetry) each in comparison with the approximate $10^{23}$ determinants for the FCI calculation, and these results are hence further evidence of the extraordinary nature of the electron correlation present in Cr$_2$~\cite{lehtola_head_gordon_fci_decomp_jcp_2017}. In terms of absolute accuracy, the MBE-FCI deviations from DMRG amount to $-1.55$ m$E_{\text{H}}$ and $2.24$ m$E_{\text{H}}$ using RHF MOs and CCSD NOs, respectively, which may be lowered to $-0.18$ m$E_{\text{H}}$ by using NOs in combination with a tighter screening threshold ($a=2.5$), that is, well within the DMRG error bar. In contrast, CCSD(T), CCSDT, and CCSDTQ differ by $22.56$ m$E_{\text{H}}$, $41.29$ m$E_{\text{H}}$, and $14.54$ m$E_{\text{H}}$, respectively, unextrapolated DMRG using a large bond dimension of 8000 by $1.45$ m$E_{\text{H}}$~\cite{chan_dmrg_2015}, and state-of-the-art perturbed heath-bath CI by $0.74$ m$E_{\text{H}}$~\cite{holmes_umrigar_heat_bath_ci_jctc_2016}.

\begin{figure}[ht]
\begin{center}
\includegraphics[scale=0.8]{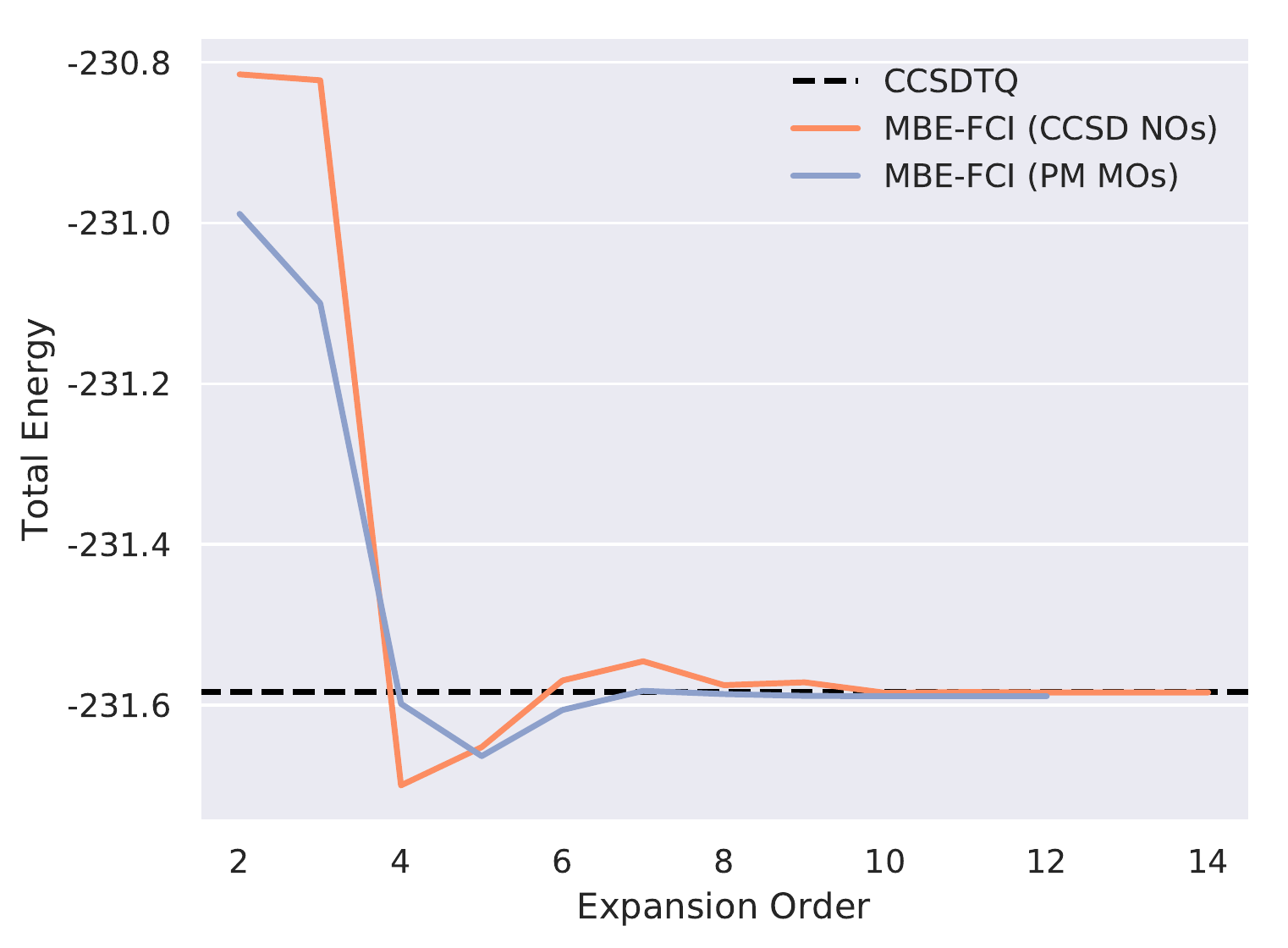}
\caption{{\textbf{Type--3}} MBE-FCI for C$_6$H$_6$/cc-pVDZ at equilibrium geometry. Results in $E_{\text{H}}$.}
\label{c6h6_dz_fig}
\end{center}
\vspace{-0.6cm}
\end{figure}
We end this Letter in Figure \ref{c6h6_dz_fig} by presenting {\textbf{Type--3}} MBE-FCI results for the benzene molecule (C$_6$H$_6$) using a cc-pVDZ basis set~\cite{dunning_1_orig} and the geometry of Ref. \citenum{sauer_thiel_cc3_benchmark_jcp_2008}. Despite being perhaps the most prominent example of a delocalized $\pi$-system, given the spatial extent across the benzene ring, we once again explore the possibility of using localized PM orbitals in addition to CCSD NOs, while we this time freeze the core MOs and correlate only the 30 valence electrons among the corresponding 108 valence and virtual MOs. As for Cr$_2$ in Figure \ref{cr2_dz_fig}, the correlation space is much too comprehensive for any exact reference energy to be obtained and we thus compare our results to high-level coupled cluster with up to quadruple excitations~\cite{ccsdtq_paper_1_jcp_1991,ccsdtq_paper_2_jcp_1992,matthews_stanton_ccsdtq_jcp_2015} (CCSDTQ) instead, as calculated using the {\texttt{NCC}} module~\cite{ncc} of the {\texttt{CFOUR}} quantum chemical program package~\cite{cfour}. In comparison with the Cr$_2$ results in Figure \ref{cr2_dz_fig}, the C$_6$H$_6$ results in Figure \ref{c6h6_dz_fig} are less irregular and both expansions are seen to qualitatively agree on a final correlation energy. In particular, the expansion in the basis of localized orbitals exhibits a very smooth and rapid convergence, yielding a correlation energy well below the CCSDTQ result ($-4.77$ m$E_{\text{H}}$), even when using an aggressive screening threshold ($a=5.0$). In comparison, the results of the expansion in the basis of CCSD NOs also falls below the CCSDTQ result by $-0.11$ m$E_{\text{H}}$, while traditional CCSD(T) and CCSDT lie $2.86$ m$E_{\text{H}}$ and $2.47$ m$E_{\text{H}}$ above CCSDTQ, respectively. In addition, Blunt, Thom, and Scott recently reported an $i$-FCIQMC result (augmented by a second-order perturbation correction) that differs by $1.69\pm0.7$ m$E_{\text{H}}$ from CCSDTQ~\cite{blunt_thom_scott_i_fcimqc_c6h6_jctc_2019} (error bars give the statistical uncertainty, not the remaining initiator error of the $i$-FCIQMC calculation). We are currently investing efforts into a more rigorous benchmark study of benzene, but for now it suffices to note that the largest CASCI calculations at the final order of the expansion using PM MOs without point group symmetry comprised only $8.54\cdot10^{5}$ determinants whereas the Hilbert space for a full C$_6$H$_6$/cc-pVDZ frozen-core FCI calculation involves in excess of an astonishing $10^{35}$ determinants.

In summary, we have presented a new powerful generalization of the recently introduced MBE-FCI method. Starting from an empty vacuum reference space, we have presented near-exact correlation energies for electron-rich model and molecular systems that would not previously have been amenable to a treatment by earlier incarnations of the method. Through results for the Hubbard model defined on a 1-dimensional lattice with between 6 and 46 sites, the challenging chromium dimer, and the ubiquitous benzene molecule---all treated on an equal and unbiased footing---we have shown that the MBE-FCI method offers a promising, intuitive, and scalable take on the electron correlation problem. While further refinements to, e.g., our screening protocol are warranted, given its extended application range, MBE-FCI now comprises a mature and versatile computational method ready to be utilized in the context of complex and intriguing problems of applied chemical interest.
 
%
%
\section*{Acknowledgments}

We are indebted to Prof. Sandeep Sharma and Ankit Mahajan of the University of Colorado Boulder for assisting us with DMRG results for the 1-dimensional Hubbard model. J. J. E. is further grateful to Prof. Frederick R. Manby of the University of Bristol for hosting his current postdoctoral fellowship and for various insightful discussions related to the present work. J. J. E. is grateful to both the Alexander von Humboldt Foundation and the Independent Research Fund Denmark for financial support. Finally, J. J. E. and J. G. acknowledge PRACE for awarding us access to Galileo at CINECA (Italy) through the 18th PRACE Project Access Call.

%
%
\section*{Supporting Information}

All results of the present work are collected in the Supporting Information as Tables S1--S4 and Figures S1--S2.

\newpage

\providecommand{\latin}[1]{#1}
\makeatletter
\providecommand{\doi}
  {\begingroup\let\do\@makeother\dospecials
  \catcode`\{=1 \catcode`\}=2 \doi@aux}
\providecommand{\doi@aux}[1]{\endgroup\texttt{#1}}
\makeatother
\providecommand*\mcitethebibliography{\thebibliography}
\csname @ifundefined\endcsname{endmcitethebibliography}
  {\let\endmcitethebibliography\endthebibliography}{}


\begin{mcitethebibliography}{57}
\bibitem[Not()]{Note-1}
The localized orbitals for the 1-dimensional Hubbard model have been computed
  by redefining the population tensor in the standard Pipek-Mezey localization
  procedure into a site-population tensor for each orbital-pair density\relax
\mciteBstWouldAddEndPuncttrue
\mciteSetBstMidEndSepPunct{\mcitedefaultmidpunct}
{\mcitedefaultendpunct}{\mcitedefaultseppunct}\relax
\EndOfBibitem
\end{mcitethebibliography}
\end{document}